\documentclass[a4paper,twoside]{article}
\usepackage{etex}
\usepackage[T1]{fontenc}
\usepackage[latin1]{inputenc}
\usepackage[english]{babel}
\usepackage{graphicx}
\graphicspath{{figures/}} 
\usepackage[noindentafter,pagestyles]{titlesec}
\usepackage{textcomp}
\usepackage{dcolumn}
\usepackage{amsmath}
\usepackage{txfonts}
\usepackage{fancyhdr}
\usepackage{threeparttable}
\usepackage{xspace}
\usepackage{pictex}
\newcommand{\bmp}[1]{\begin{minipage}{#1\columnwidth}}
\newcommand{\emp}{\end{minipage}}

\newcommand{\bea}{\begin{eqnarray}}
\newcommand{\eea}{\end{eqnarray}}
\newcommand{\be}{\begin{equation}}
\newcommand{\ee}{\end{equation}}

\titleformat{\section}{\large\bfseries}{\thesection.}{.3em}{}
\titlespacing*{\section}{\leftmargini}{*3}{*3}
\titleformat{\subsection}{\bfseries}{\thesubsection}{.3em}{}
\titlespacing*{\subsection}{0pt}{*3}{*3}
\makeatletter
\def\@maketitle{%
  \newpage
  \null
  \vskip 2em%
  \begin{center}%
  \let \footnote \thanks
    {\fontsize{18}{22}\fontseries{b}\selectfont \@title \par}%
    \vskip 1.5em%
    {\normalsize
      \lineskip .5em%
      \begin{tabular}[t]{c}%
\@author
      \end{tabular}\par}%
    \vskip 1em%
    {\large \@date}%
  \end{center}%
  \par
  \vskip 1.5em}
\renewenvironment{abstract}{%
\if@twocolumn
\section*{\abstractname}%
\else
\quotation
\noindent{\bfseries\large \abstractname\vspace*{.3ex}\par}
\fi}
{\if@twocolumn\else\endquotation\fi}
\makeatother

\fancypagestyle{plain}{%
} 
\usepackage{algpseudocode}
\usepackage{algorithm}
\usepackage{pgfplots}
\usepgfplotslibrary{statistics}
\usepackage{placeins}
\usepackage{subcaption}
\usepackage{xcolor}
\usepackage[short]{optidef}
\newcommand{\tp}{{\mkern-1.5mu\mathsf{T}}} 
\begin{document}
%
%
%
%
%
\title{Sequential Convex Programming for Optimal Line of Sight Steering in Agile Missions}
\author{\itshape J.\ Ram\'irez$^\star$ and L.\ Hewing$^{\star\dag}$ \\
\itshape$^\star$SENER Aerospace\\
\itshape Tres Cantos, Madrid, Spain \\
{[jesus.ramirez\,$\vert$\,lukas.hewing]}@aeroespacial.sener\\
$^\dag$Corresponding author}

\date{}
\maketitle
\begin{abstract}
  \noindent 
The trend toward onboard autonomy and spacecraft minimization present significant potential for advances in efficient Line of Sight management by making optimal use of the limited torque resources available. At SENER Aeroespacial, we are implementing AOCS algorithms capable of providing agility in different observation scenarios in which the exploitation of the resources is fundamental for the mission success. In this contribution, we present an in-house optimization toolbox for onboard guidance, the SENER Optimization Toolbox, and we propose its use for online attitude guidance of an agile spacecraft using Control Moment Gyroscopes. We propose different optimization schemes based on Sequential Convex Programming aiming for reducing the computational burden for real-time implementations. The results highlight the potential for performance improvements when making use of embedded optimal control for fast slew maneuvers: with proposed schemes, we find solutions that implicitly manage singularities and are up to $12.2\%$ faster than classical \textit{bang-bang} maneuvers while increasing the smoothness of the trajectories to minimize the excitation of flexible modes.
\end{abstract}

\section{Introduction}
There is a growing demand for ever-larger amounts of high-quality data within the Earth observation market~\cite{EO_Survey}. One way to improve image quality is to enhance the performance of the optical payloads, be it through improvements at the payload level itself, or at AOCS levels via improved pointing stability. Increasing the quantity of available information, on the other hand, requires increased access
rates to geographical locations of interest, as well as enhanced coverage and revisit times, thus calling for agile maneuvers. Together with the current trend toward onboard autonomy and spacecraft miniaturization, this presents significant potential for advances in efficient Line of Sight (LoS) steering by making optimal use of the limited resources available~\cite{Florio2006}.

Commonly, spacecraft attitude acquisition is a slow process that is sequenced in safe maneuvers at well-known few moments of a mission, in comparison with the total duration. For the rest of the spacecraft's life, the acquired attitude, e.g., nadir pointing, is maintained by controlling the error with, e.g., a proportional derivative control, without needing further guidance. The required slew maneuvers are specified in the mission planning, and thanks to lax requirements in time for completing them, they can be easily designed with conservatism for ensuring a safe operation, e.g., not pointing sensitive payload towards bright objects, not losing communications, or not saturating the control devices.

When time plays a central role, as it does in \textit{Agile Earth Observation Satellites} missions, these slow sequenced slews are no longer valid. So-called agile maneuvers -- large attitude re-configurations at high speeds -- present large challenges on mechanical, system, and GNC level. For instance, Geo-Eye 1 can provide 1 deg/s turning rates~\cite{GEO1}, while missions like Pleiades~\cite{Thieuw2007} or SPOT 6 and 7~\cite{Spot67} already achieve rates of 2-3 deg/s. This has increased relevance under the use of Control Moment Gyroscopes (CMGs)
with complex momentum envelopes, which are increasingly employed in the field of agile missions.

As an alternative to classical approaches, we propose to use embedded numerical optimal control for attitude guidance in interaction with a target scheduler. This way, we maximize observation profit by optimally exploiting the
torque resources of the spacecraft while minimizing the time between targets. Furthermore, with this approach, we reduce operational costs related to mission planning. This results in a challenging constrained nonlinear optimization problem that, thanks to the advances in computational resources and convex optimization, becomes increasingly tractable for onboard implementation. Along this line, we have developed the SENER Optimization ToolBox (SOTB), an autocodable MATLAB implementation of a toolbox for numerical optimal control which is independent of other MATLAB toolboxes or external software. It includes a Sequential Convex Programming (SCP) algorithm powered by an interior-point solver, both developed in-house.

This paper outlines the current state of SOTB and presents a demonstrator application in form of a small Earth observation spacecraft with CMGs that is to be reconfigured
to point to a new target. Depending on the specific formulation at hand, we express actuator constraints via polytopes or ellipsoids and generally obtain the overall solution by solving a sequence of quadratically-constrained quadratic programs. We  focus on minimizing the time of the maneuver, for which propose two different formulations. The first consists of a direct optimization from the lowest level (gimbal rates) to the spacecraft attitude, while the second is based on an iterative sequence of two linked optimization problems: the attitude guidance problem (from torque to attitude), and the dynamic CMG allocation (from gimbal rates to momentum).

The paper is organized as follows. Section~\ref{sc:SOTB} outlines the SOTB, including its interior-point solver for convex quadratically constrained quadratic problems (QCQPs) and an SCP algorithm focusing on optimal control problems. Section~\ref{sc:cmgCaseStudy} then presents the SOTB application case to a CMG actuated attitude guidance problem while Section~\ref{sc:conclusion} contains a conclusion and short discussion.

\section{SENER Optimization ToolBox (SOTB)}\label{sc:SOTB}
The SOTB contains functions for general numerical optimization with a specific focus on optimal control, both linear and non-linear using Sequential Convex Programming (SCP) schemes. The current state of the development allows the research and quick prototyping of optimal control problems and is being tested in industrial applications. The following sections give a brief overview of the interior-point solver for quadratically constrained quadratic programs (QCQP), as well as the Sequential Convex Programming (SCP) tool which is powered by the first. 

\subsection{Quadratically Constrained Quadratic Programming (QCQP)}
As a central tool the SOTB provides a solver for convex quadratically constrained quadratic programs (QCQP), i.e., optimization problems of the form:
\begin{mini!}{x \in \mathbb{R}^n}{\frac{1}{2} x^\tp H x +c^\tp x}{\label{eq:QCQP}}{} \label{Eq1a}\
\addConstraint{A_{\text{eq}}x = b_{\text{eq}}} \label{Eq1b}
\addConstraint{A_{\text{ineq}}x \leq b_{\text{ineq}}}\label{Eq1c}
\addConstraint{\frac{1}{2} x^\tp Q_i x + q_i^\tp x \leq 1 \ \forall i=1,...n_q} \label{Eq1d}
\end{mini!}
where the problem is \emph{convex}, i.e., the matrices of the quadratic forms are positive semidefinite $H \succeq 0$, $Q_i \succeq 0$. The matrices $A_{\text{eq}}$, $ A_{\text{ineq}}$, and vectors $b_{\text{ineq}}$, $b_{\text{ineq}}$ are used to express affine equality and inequality constraints, i.e. polytopic constraints. For convenience, the quadratic forms in \eqref{Eq1a} and \eqref{Eq1d} can be expressed w.r.t.\ their center, e.g.\ $x-x^c_i$.

The solver is based on a primal-dual interior-point method (IPM)~\cite{Nocedal2006, Gros2020} which aims to solve for the first-order optimally conditions of \eqref{eq:QCQP}, that is the Karush-Kuhn-Tucker (KKT) conditions. Given suitable constraint qualification, these conditions are necessary and sufficient for optimality and can be expressed as
\begin{equation}\label{eq:KKT}
\begin{bmatrix} H  & 0 & \nabla g(x)^\tp  & A_{\text{eq}}^\tp \\ 
A_{\text{eq}} & 0 & 0 & 0 \\
\nabla g(x) & I & 0 & 0 \\ 
0 & 0 & S & 0 \end{bmatrix} \begin{bmatrix} x \\ s \\ \mu \\ \lambda \end{bmatrix} = \begin{bmatrix} -c \\ b_{\text{eq}} \\ \bar{b}_{\text{ineq}} \\ 0 \end{bmatrix}
\end{equation}
in which $s \geq 0$ is a vector of slack variable, $S = \text{diag}(s)$, $\mu$ and $\lambda$ are the Lagrange multipliers of the inequality and equality constraints, respectively, and $\nabla g(x) = [A^\tp_{\text{ineq}}, \,   Q_1x + q_1, \, \ldots, \,  Q_{n_q}x + q_{n_q}]^\tp$, $\bar{b}_{\text{ineq}} = [b_{\text{ineq}}^\tp, \textbf{1}^\tp]^\tp$. Equation~\eqref{eq:KKT} is nonlinear due to $\nabla g(x)$ and in particular the complementary slackness condition $S\mu = 0$ making direct solutions extremely challenging. In IPM, the problem is relaxed using a smoothing constant or barrier parameter $\tau > 0$ on this complementarity condition as $S\mu = \tau \textbf{1}$, leading to the following Newton search direction
\begin{equation}\label{eq:KKT_Newton}
\begin{bmatrix} H +\sum_{i=1}^{n_q}\mu_i Q_i  & 0 & \nabla g(x)^\tp  & A_{\text{eq}}^\tp \\ 
A_{\text{eq}} & 0 & 0 & 0 \\
\nabla g(x) & I & 0 & 0 \\ 
0 & Z & S & 0 \end{bmatrix} \begin{bmatrix} \Delta x \\ \Delta s \\ \Delta \mu \\ \Delta \lambda \end{bmatrix} = \begin{bmatrix} -c  -Hx - \nabla g(x)^\tp \mu - A_{\text{eq}}^\tp \lambda \\ b_{\text{eq}} - A_{\text{eq}} x \\ \bar{b}_{\text{ineq}} - \nabla g(x) x - s \\ S \mu - \tau \textbf{1} \end{bmatrix}
\end{equation}
where $Z = \text{diag}(\mu)$. System \eqref{eq:KKT_Newton}
 is solved for successively decreasing values of $\tau$ and using a line search procedure in the corresponding search direction to obtain the solution to the original problem~\eqref{eq:QCQP}.
Summarizing the methodology implemented in the SOTB, some of the main characteristics are: 
\begin{itemize}
  \setlength\itemsep{0.1em}
    \item Use of slack variables $s$ for inequality constraints, enhancing simple treatment of allowed unfeasible initial guess. 
    \item Line-search maintaining feasibility in the relaxed problem. 
    \item The Newton step \eqref{eq:KKT_Newton} is computed following a \textit{corrector and centering path} methodology~\cite{Nocedal2006} consisting of \emph{(a)} an \textit{affine} direction calculation (solution of \eqref{eq:KKT_Newton} with $\tau = 0$), \emph{(b)} a barrier parameter affine-scaling and \emph{(c)} the solution of the resulting KKT augmented system \eqref{eq:KKT_Newton}. 
\end{itemize}

\paragraph{Solver Performance}
The solver has been tested against \texttt{MATLAB} \texttt{@quadprog} solver with its default \textit{interior-point} algorithm in a benchmark collection from \textit{qpOases}~\cite{QPcollection}, demonstrating stability and competitive performance. 
The SENER tool is able to solve ill-conditioned problems of the collection like \textit{crane} or \textit{robotic-arm}, in which \texttt{@quadprog} fails. Table~\ref{Table_QCQP_benchmark} shows some performance metrics.

\begin{table}
\centering
\caption{Benchmark QP performance comparison of SOTB against \texttt{$@$quadprog}}     \label{Table_QCQP_benchmark}
\begin{tabular}{ccccc}
         \bf QP size& \multicolumn{2}{c}{ \bf SOPT QCQP}  & \multicolumn{2}{c}{ \bf \texttt{$@$quadprog}} \\
          (\# variables | \# constraints)& Median \# iterations & Mean solve time [s] & Median \# iterations & Mean solve time [s] \\ \hline
          2\,|\,4&5&0.002&4 &0.003\\
          256\,|\,1154&20&0.143&22 &0.662\\
          240\,|\,480&8&0.092& 6&0.006\\
          20\,|\,80&9&0.010&8 &0.003\\
          60\,|\,630&117&0.438&\textit{Not solved}&\textit{Not solved}\\
          75\,|\,470&37&0.154&\textit{Not solved}&\textit{Not solved}\\
          57\,|\,434&35&0.120&\textit{Not solved}&\textit{Not solved}\\
\hline
\end{tabular}
\end{table}

\subsection{Sequential Convex Programming (SCP)}
The SOTB includes a sequential convex programming (SCP) tool making use of the previously described QCQP solver. It is developed with a focus on numerical optimal control of nonlinear systems, in particular for guidance functionality. The optimization problems addressed by the SCP tool have the form:
\begin{mini!}{x_i \in \mathbb{R}^{n_x},u_i \in \mathbb{R}^{n_u}, \gamma \in \mathbb{R}^{n_s}}{\frac{1}{2} x_{N}^TPx_{N}+p^Tx_{N}+
\frac{1}{2}\gamma^TC\gamma+c^T\gamma+
\sum_{i=0}^{N-1}  \frac{1}{2}x_i^TL_ix_i+l_i^Tx_i+
\frac{1}{2}u_i^TR_iu_i+r_i^Tu_i}{\label{eq:SCP}}{}\label{Eq3a}
\addConstraint{x_{i+1}=f_i(x_i,u_i);\quad \forall i=0,..,N-1}\label{Eq3b}
\addConstraint{A_i^xx_i\leq b^x_i+J^{\text{aff},x}_i\gamma;\quad \forall i=1,...,N}\label{Eq3c}
\addConstraint{A_i^uu_i\leq b^u_i+J^{\text{aff},u}_i\gamma;\quad \forall i=0,...,N-1}\label{Eq3d}
\addConstraint{\frac{1}{2}x_i^TQ_{i,j}^xx_i+(q_{i,j}^x)^Tx_i\leq c_{i,j}^x+(J_{i,j}^{q,x})^T\gamma; \quad \forall j = 1,...,n_{xi}^q; \forall i= 1,...,N}\label{Eq3e}
\addConstraint{\frac{1}{2}u_i^TQ_{i,j}^uu_i+(q_{i,j}^u)^Tu_i\leq c_{i,j}^u+(J_{i,j}^{q,u})^T\gamma; \quad \forall j = 1,...,n_{ui}^q; \forall i= 0,...,N-1}\label{Eq3f}
\addConstraint{\gamma\geq 0}\label{Eq3g}
\addConstraint{x_0 = x_{\text{init}}}\label{Eq3h}
\end{mini!}
where $x_i$ is the $n_x$ discrete system state at time step $i$, $u_i$ the $n_u$ control inputs, and $\gamma$ a vector of $n_s$ slack variables. The problem, therefore, follows the typical stage structure of optimal control with $N$ being the control horizon and the dynamics constraint~\eqref{Eq3b} coupling the different stages. The problem is subject to a set of polytopic (\ref{Eq3c})-(\ref{Eq3d}) and quadratic (\ref{Eq3e})-(\ref{Eq3f}) constraints that can be softened with slack variables $\gamma$. The slack variables are defined globally, that is outside the stage structure, and selected stage-wise for the different constraints using the matrices $J_i^{(\cdot)}$. In the current implementation, the cost function is convex quadratic, making the dynamics the only source of nonlinearity. This pre-defined structure allows for some extra modifications by the user. For example, constraints relating different stages -- for instance rate constraints on the input variables -- or combined state and control constraints are possible.
 As input, the SCP tool requires the dynamics functions $f_i(x_i,u_i)$ as well as its first-order Taylor expansion, i.e.\ its derivatives with $x$ and $u$. To facilitate this,  the SOTB includes a tool to discretize time-invariant continuous-time dynamics $\dot{x}=f_c(x,u)$ to $x_{i+1}=f(x_i,u_i)$, e.g.\ following a Runge-Kutta 4 scheme under zero-order-hold.

The algorithm implemented follows a common SCP logic, solving iteratively a sub-problem of the original until a convergence criterion is matched, see e.g.\ \cite{Malyuta2021, Messerer2021, Samir2021}. The main steps in translating \eqref{eq:SCP} into a form amenable to the QCQP solver are
\begin{enumerate}
    \item The nonlinear dynamics \eqref{Eq3b} are linearized around the solution of the previous iteration $({\bar{x}_i},{\bar{u}_i})$, or from an initial guess. This changes \eqref{Eq3b} to
    \begin{equation}
        x_{i+1}=\nabla_x f_i({\bar{x}_i},{\bar{u}_i})(x_i-\bar{x}_i)+\nabla_u f_i({\bar{x}_i},{\bar{u}_i})(u_i-\bar{u}_i)+f_i({\bar{x}_i},{\bar{u}_i});\quad \forall i=0,..,N-1\label{Eq3b2}
    \end{equation}
    \item Variable quadratic trust-region constraints are included to aid convergence of the SCP
    \begin{subequations}
    \begin{align}
        \frac{1}{2} \Vert x_i-\bar{x}_i\Vert^2 &\leq \delta^X_{\max} \  \forall i= 1,...,N \, ,\label{Eq3i} \\ 
        \frac{1}{2}\Vert u_i-\bar{u}_i \Vert^2 &\leq \delta^U_{\max} \ \forall i= 0,...,N-1 \, .\label{Eq3j}
    \end{align}
    \end{subequations}
    Here, the trust-region parameters $\delta^X_{\max}$, $\delta^U_{\max}$ are adjusted dynamically based on the optimization progress \cite{Malyuta2021}.
\end{enumerate}
Algorithm~\ref{alg:sotbSCP} outlines the SCP scheme implemented in the SOTB. 
\begin{algorithm}
\caption{sotbSCP pseudo-algorithm} \label{alg:sotbSCP}
\begin{algorithmic}
\State \textbf{Input:}
\State Problem parameters in~\eqref{eq:SCP} including $f_i$ and $\nabla f_i$
\State Iteration limits $N_{\text{iter}}$ and tolerances $\epsilon$, $\epsilon_{\text{TR}}$
\State Initial trajectory guess $(X_0,U_0)$;
\State Initial values for trust-region $\left(\delta^X_{max},\delta^U_{max},\kappa^+,\kappa^-\right)$;

\State
\State \textbf{Sequential Convex Programming:}
\State $z = (X_0,Y_0,1)$;
\State $(A_{\text{eq}},b_{\text{eq}}) \gets$ Linearization of $f_k$ around $z$;
\State $\delta_{\text{prog}} = \infty$
\While{$\delta_{\text{prog}} \geq \epsilon$}
\State $\bar{z} \leftarrow z$
\State $\delta_{\text{TR}} = \infty$
\While{$\delta_{\text{TR}} \geq \epsilon_{\text{TR}}$}
\State $z \gets $ Minimizer of optimization sub-problem with
$\left(A_{\text{eq}},b_{\text{eq}},\bar{z},\delta^X_{\max},\delta^U_{\max}\right)$;
\State $(A_{\text{eq}}^*,b^*_{\text{eq}}) \gets$ Linearization of $f_k$ around $z$;
\State $\delta_{\text{TR}} \gets ||A_{\text{eq}}^*z-b_{\text{eq}}^*||_2 $
\If{$\delta_{\text{TR}} \leq \epsilon_{\text{TR}}$} 
	\State $\left(\delta^X_{max},\delta^U_{max}\right) \gets \left(\kappa^+\delta^X_{max},\kappa^+\delta^U_{max}\right)$;
\Else 
	\State $\left(\delta^X_{max},\delta^U_{max}\right) \gets \left(\kappa^-\delta^X_{max},\kappa^-\delta^U_{max}\right)$;	    
\EndIf 
\EndWhile
	\State $(A_{eq},b_{eq}) \gets (A_{eq}^*,b_{eq}^*)$;    
	\State $\delta_{\text{prog}} \gets ||z-\bar{z}||_2$
\EndWhile

\State \textbf{Output:} Minimizer $z$
\end{algorithmic}
\end{algorithm}

\section{Application case: Spacecraft agile slews with CMG cluster}\label{sc:cmgCaseStudy}
In this section, we are going to outline some of the work at SENER Aeroespacial in the field of agile missions, applying the  developed sequential convex programming tool to a challenging attitude control problem under control moment gyroscope actuation. In particular, we consider a simple attitude slew, a rotation around a single axis of $\Delta \theta$ degrees, optimizing jointly over the attitude and the individual CMG gimbal trajectories.

For comparison, we consider a bang-bang-like torque profile considering the maximum instantaneous torque capabilities of the CMGs available in the demanded direction. This solution is known to be (time) sub-optimal for the attitude slews around non-principal axes \cite{Wertz,Scrivener1994}, which is further complicated by the complex CMG trajectories and potential singularities \cite{Leve}.
We show that numerical optimal control solutions have the potential to significantly reduce maneuver time while increasing the smoothness of the trajectory while exploiting the CMG cluster capabilities and managing its singularities. 

\subsection{Dynamics formulation}
The following presents the dynamics used for the simulations studies, both in terms of the spacecraft attitude, see Section~\ref{ssc:attitudeDynamics}, and the CMG actuation, see Section~\ref{ssc:CMGDyanmics}. All parameters used are summarized in Table~\ref{Table_SC_properties}, which are taken from a similar study in~\cite{Geshnizjani2020}. 

\begin{table}
\centering
\caption{Spacecraft properties for the application case numerical results}     \label{Table_SC_properties}
\begin{tabular}{lcllcl}
\bf Quantity& \bf Value  & \bf Unit & \bf Quantity& \bf Value  & \bf Unit \\ \hline 
Moments of Inertia $\left[ J_{xx},J_{yy},J_{zz}\right]$ & $\left[ 5000,5000,3000\right]$ & kgm$^2$ & Roof tip angle $\beta$ & 45& deg\\
Products of Inertia $\left[ J_{xy},J_{xz},J_{yz}\right]$ & $\left[ -400,-70,-80\right]$ & kgm$^2$ & Maximum gimbal rate $\dot{\delta}_{max}$ & 1 & rad/s\\
DCM cluster to body frame & $T_x\left(\frac{pi}{6}\right)T_z\left(\frac{pi}{2}\right)T_x\left(\frac{pi}{2}-\beta\right)$ & &CMG momentum $h_{CMG}$ & 100 & Nms  \\
\hline
\end{tabular}
\end{table}
\subsubsection{Attitude}\label{ssc:attitudeDynamics}
The motion of a rigid spacecraft in an inertial reference frame controlled by a momentum device can be expressed as 
\begin{equation}
   \dot{h} = J\dot{\omega}=\tau \label{Eq4d2}
\end{equation}
in which $h$ is the spacecraft momentum, $J$ is the spacecraft inertia in body frame, $\omega$ the inertial angular rate expressed in the body frame, and $\tau$ the applied torque of the momentum device. For simplicity, we assume here that no external torques are applied to the spacecraft and that we have no residual momentum, i.e. we assume that a suitable offloading strategy is used\footnote{Note that this is largely for simplicity of the exposition, the presented optimization approaches can be similarly applied to more complex configurations, albeit under a potential slight increase of computational complexity}. Integration of the angular rates can then be done via

\begin{equation}
   \begin{bmatrix} \dot{\phi} \\ \dot{\theta} \\ \dot{\psi} \end{bmatrix} = \begin{bmatrix} (\omega_x\cos{\psi}-\omega_y\sin{\psi})\text{sec}\theta \\
    \omega_x\sin{\psi}+\omega_y\cos{\psi}\\ 
   \omega_z - (\omega_x\cos{\psi}-\omega_y\sin{\psi})\text{tan}\theta
   \end{bmatrix} \, ,
   \label{Eq4}
\end{equation}
where the spacecraft attitude is represented by Euler angles \emph{roll} $\phi$, \emph{pitch} $\theta$, and \emph{yaw} $\psi$ with respect to the inertial frame. Note that due to the assumption of zero residual momentum and non-external forces, the current spacecraft momentum is always opposite and equal to the stored momentum in the momentum device.

\subsubsection{Control Momentum Gyroscope}\label{ssc:CMGDyanmics}
\begin{figure}
\begin{subfigure}{0.45\textwidth}
\hfill
\includegraphics[height=60mm]{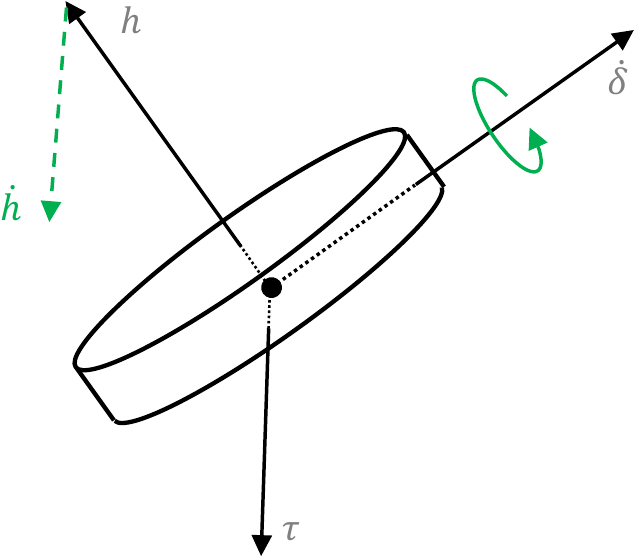}
\caption{CMG torque generation via the gyroscopic principle}
\label{CMGscheme}
\end{subfigure} 
\hfill
\begin{subfigure}{0.45\textwidth}
\centering
\includegraphics[height=60mm]{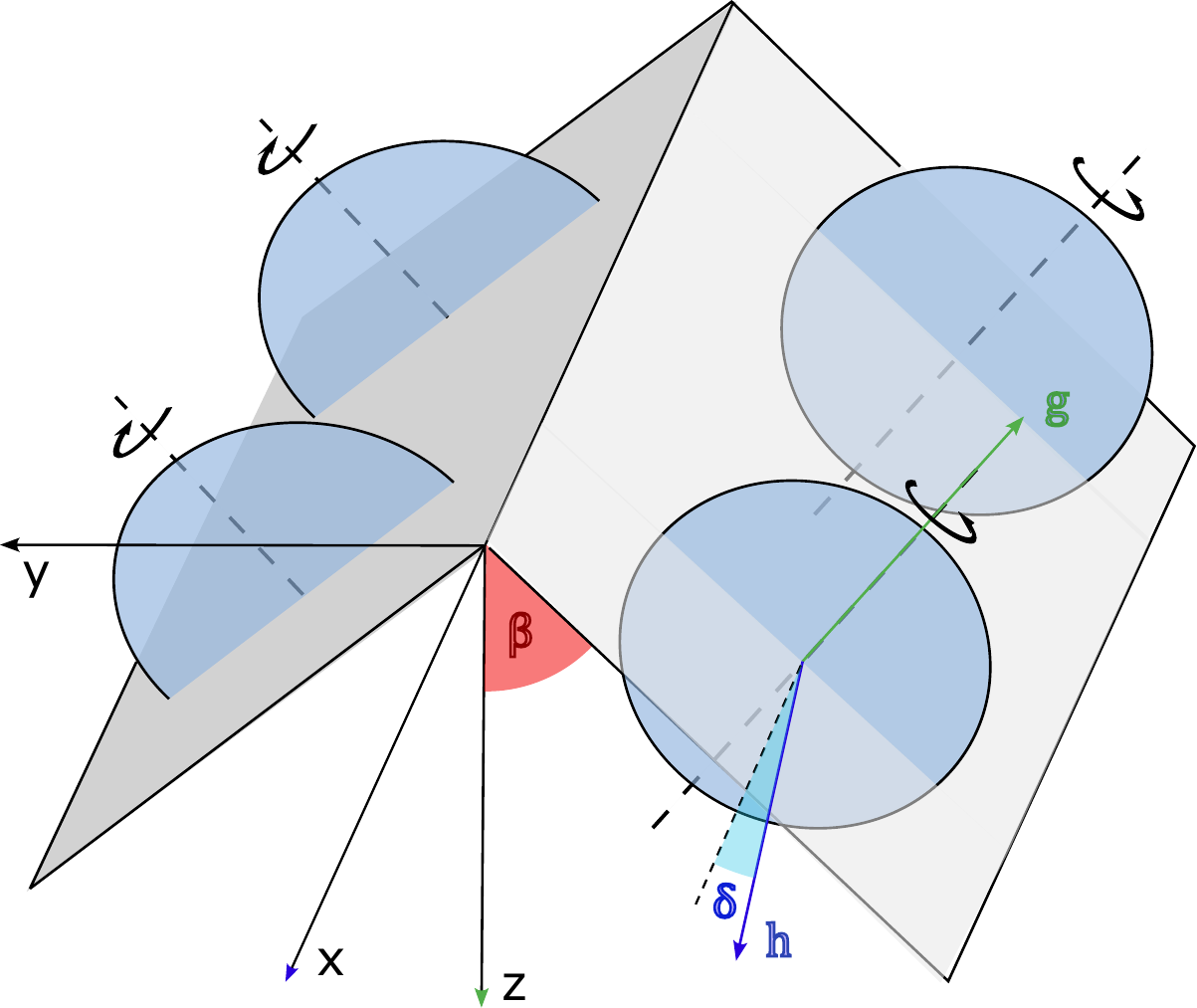}
\caption{Roof CMG cluster scheme}
\label{ClusterDef}
\end{subfigure}
\hfill
\caption{CMG illustrations}
\end{figure}

As momentum devices, we consider an array of constant speed single gimbal axis CMGs, which are particularly torque-efficient with respect to their mass and therefore of interest in agile missions of small-scale satellites. These actuators operate by changing the axis of rotation of a spinning flywheel which results in a perpendicular torque. Figure~\ref{CMGscheme} presents an illustration of this, in which $h_i$ is the current momentum of the CMG, $\dot{\delta}_i$ the gimbal rate -- which we take as the control input to the system -- and $\tau_i$ the resulting torque perpendicular to the current momentum and gimbal axis. Similar to reaction wheels, CMGs are typically operated in clusters or arrays in order to allow the generation of torque in all spatial dimensions. Neglecting lower-order effects \cite{Leve}, the torque provided by a CMG cluster can be expressed as
\begin{equation}
  \tau= - \sum_{i=1}^{n_{CMG}} \hat{g}_i \times \hat{h}_i(\delta_i) \dot{\delta}_i = -C(\delta) \dot{\delta} \label{Eq4f}
\end{equation}
in which $\delta_i$ is the scalar gimbal rate of CMG $i$, $\hat{g}_i$ is the unit vector describing the gimbal axis, and $\hat{h}_i(\delta_i) = \frac{h_i(\delta_i) }{||\mathbf{h}_i(\delta_i) ||_2}$ is the direction of the CMG momentum vector which is a function of the current gimbal angle $\delta_i$. This relationship can be summarized using the so-called Jacobian $C(\delta)$ and the vector of gimbal angles $\delta$. The torque capabilities of a CMG array are then typically limited by a maximum gimbal rate of each CMG in the array, which we consider as a constraint of the form
\begin{equation}
    \left|{\dot{\delta}}_{i}\right|\leq {\dot{\delta}}_{{\max}} 
\end{equation}

Complexities in the control of CMG arrays can now arise from the fact that the direction of torque that a single CMG can produce instantaneously changes over time as it is aligned with the current momentum vector $h_i$. In particular, there exist CMG angles $\delta$ (gimbal configurations) for which the Jacobian $C(\delta)$ loses rank, meaning that torque can only be generated on a subspace of $\mathbb{R}^3$, i.e.\ a plane or a line. These configurations are called \emph{singular} and are particular to the specific accommodation of CMGs at hand, making the singularity study a complex and challenging topic~\cite{Leve}. 

For these studies, we have selected a roof pyramid of tip angle $\beta$. With $\gamma_i$ specifying the cant of the plane whose normal is the gimbal axis of the $i$-th CMG, see Figure \ref{ClusterDef} \cite{Leve}. For this configuration, we can express the momentum direction as
\begin{equation}
   {\hat{h}}_{i}(\delta_i)=\left[\begin{matrix}\cos(\delta_i)\cos(\beta)\cos(\gamma_i)-\sin(\delta_i)\sin(\gamma_i)\\
   \cos(\delta_i)\cos(\beta)\sin(\gamma_i)+\sin(\delta_i)\cos(\gamma_i)\\
   -\cos(\delta_i)\sin(\beta) \end{matrix}\right] \label{Eq4h} \, .
\end{equation}
Note, however, that it is straightforward to address different configurations of CMGs.
\subsection{Baseline Guidance Solution: Bang-Bang}\label{ssc:Baseline}
For comparison purposes, we compute \textit{bang-bang}-like torque profiles for performing a single \emph{eigenaxis rotation}~\cite{Wertz}. These profiles are characterized for delivering maximum torque in a fixed direction $\hat{\tau}$ of the eigenaxis rotation until a certain velocity is reached, followed by a free drifting motion period and opposite maximum torque until the motion is stopped. 
In our case, we generate the CMG actuation $\dot{\delta}$ by computing a gimbal rate direction via the \textit{Moore-Penrose Pseudoinverse} (MPP) from the torque direction \cite{Leve} 
\begin{equation}
    \hat{\dot{\delta}} = -C(\delta)^\tp \left(C(\delta) C(\delta)^\tp\right)^{-1} \hat{\tau} 
\end{equation}
which we then scale such that the gimbal rate limits are respected
\begin{equation}
   \dot{\delta} = \frac{\dot{\delta}_{\max}}{\Vert \hat{\dot{\delta}}\Vert_\infty} \hat{\dot{\delta}}\, ,   
\end{equation}
where $\Vert \cdot \Vert_\infty$ is the infinity or maximum norm which is equal to the largest absolute value of each element of the vector. We command the maximum torque in this direction until we reach $98\%$ of the achievable momentum with the cluster at which time we apply zero torque. After covering half of the commanded slew angle, we reverse the computed control inputs to come to a stop. 

Note that this baseline provides a very aggressive and time-optimized maneuver, but does not provide any singularity avoidance or escape functionality, aside from the external ones that are not reached by constraining the maximum velocity. Furthermore, the maneuver has infinite jerk, in particular also when stopping the motion which is known to lead to undesirable excitation of spacecraft modes~\cite{Kimand2006}.

\subsection{Optimal control problem formulation}

With the objective of improving the baseline solution maneuver time, but also regarding jerk, smoothness of the solution, and CMG singularity management, we propose to use numerical optimization via the SOTB to derive suitable feed-forward guidance profiles.
We distinguish between two different approaches. The first, \emph{joint optimization}, optimizes the spacecraft attitude profile directly using the CMG gimbal rates as input. In the second, \emph{sequential optimization}, we decompose the problem into two parts that are solved sequentially: An attitude guidance problem and a CMG steering problem.
Note that both make use of sequential convex programming for their respective solutions.

\paragraph{CMG Change of Variables}
Aiming for simplification and better numerical behavior in the SCP, we perform a change of variable so that
\begin{subequations}
\begin{align} 
    \alpha &= [\cos(\delta_1), \, \sin(\delta_1),...,\, \cos(\delta_4),\, \sin(\delta_4)]^\tp \\
    \dot{\alpha} &= [-\sin(\delta_1) \dot\delta_1, \, \cos(\delta_1) \dot\delta_1,\,  ...,\, -\sin(\delta_4)\dot{\delta}_4,\, \cos(\delta_4)\dot{\delta}_4]^\tp \label{AlphaChangeVble2} 
\end{align}
\end{subequations}
which allows for a linear parameterization of the spacecraft momentum which, in the considered case without residual momentum, is equal and opposite to that of the CMG cluster
\begin{equation}
     h= -C_H \alpha
\end{equation}
with $C_H$ given from~\eqref{Eq4h} as 
\begin{equation}
   C_H = h_{CMG}\begin{bmatrix}
    1 & 0 & 1 & 0 & 1 & 0 & 1 & 0 \\
    0 & \sin{\beta} & 0 & \sin{\beta}& 0 & - \sin{\beta} & 0 & - \sin{\beta}\\
    0 & \cos{\beta} & 0 & \cos{\beta}& 0 &  \cos{\beta} & 0 & \cos{\beta}
    \end{bmatrix}
\end{equation}
It also allows and a torque parameterization as
\begin{equation}
     \tau(t) = -C_\tau \Lambda(\alpha) \dot{\delta}
\end{equation}
with $\Lambda(\alpha) = \text{blkdiag}\left([\cos(\delta_1),\sin(\delta_1)]^\tp, \ldots, [\cos(\delta_4),\sin(\delta_4)]^\tp\right)$ and
\begin{equation}
   C_\tau = h_{CMG}\begin{bmatrix}
    0 & -1 & 0 & -1 & 0 & -1 & 0 & -1  \\
    \sin{\beta} & 0 & \sin{\beta}& 0 & - \sin{\beta} & 0 & - \sin{\beta} & 0\\
    \cos{\beta} & 0 & \cos{\beta}& 0 &  \cos{\beta} & 0 & \cos{\beta} & 0
    \end{bmatrix}
\end{equation}

\subsubsection{Joint optimization}
In order to formulate the optimization problem, we consider the system state $x = [\phi,\theta,\psi,\omega,\alpha]^\tp$ composed of the Euler orientation angles, the vector of angular rates $\omega$, and the $\alpha$ vector containing the gimbal angle information. As input we consider the vector of gimbal rates $u = \dot\delta$, such that the continuous-time dynamics $f_c(x,u)$ are directly given by \eqref{Eq4}, \eqref{Eq4d2} with \eqref{Eq4f}, and \eqref{AlphaChangeVble2}, resulting in 
\begin{equation}
\dot{x}=\begin{bmatrix} \begin{bmatrix}
\dot{\phi}\\
\dot{\theta}\\
\dot{\psi}
\end{bmatrix}\\
\dot{\omega}\\
\dot{\alpha}
\end{bmatrix}= f_c(x,u) =
\begin{bmatrix} \begin{bmatrix} 
(\omega_x\cos{\psi}-\omega_y\sin{\psi})\text{sec}\theta\\
\omega_x\sin{\psi}+\omega_y\cos{\psi} \\
\omega_z - (\omega_x\cos{\psi}-\omega_y\sin{\psi})\text{tan}\theta  \end{bmatrix} \\
-J^{-1}C_\tau \Lambda(\alpha) \dot{\delta}\\
 \text{Eq. \eqref{AlphaChangeVble2}} \end{bmatrix} \, .
\end{equation}
Using a discretization scheme to get $f(x,u)$ we then formulate the following nonlinear opimization problem as
\begin{mini!}{x_i ,\, u_i }{\frac{1}{2} ||x_{N}||_{L_f}^2+ ||x_N||_{S_f,\infty}
+
\sum_{i=0}^{N-1}  \frac{1}{2}||x_i||_L^2+ ||x_i||_{S,\infty}
+\frac{1}{2}||u_i\Vert_R^2}{\label{eq:joint}}{}
\addConstraint{x_{i+1}=f(x_i,u_i);\quad \forall i=0,..,N-1}
\addConstraint{|u_i|\leq \dot{\delta}_{max};\quad \forall i=0,...,N-1}
\addConstraint{x_0 = x_{\text{init}}}
\end{mini!} 

For notational convenience, we align the reference coordinate system such that the origin coincides with the target attitude, i.e. we consider a regularization problem. We make use of a mix of quadratic and infinity norm costs, which are weighted by positive semi-definite cost matrices $L_f$, $S_f$, $L$, $S$, and $R$. In particular, the cost matrices affecting the states are selected such that only affect the attitude and angular rates of the spacecraft, i.e.\ the gimbal angles remain completely free\footnote{Note that one could could include a cost on the gimbal angles, e.g. for a preferred gimbal angles approach}.
In the problem \eqref{eq:joint} we include a strong $\infty$-Norm of the attitude and velocity error at the terminal time step using $S_f$, similar to a soft constraint, see e.g.~\cite{Verschueren2019},
which enforces that the target attitude is reached at the end of the horizon with 0 velocity. Similarly, we include an infinity norm stage cost, which encourages a fast termination of the maneuver, approximating a minimum time solution to the considered problem. Quadratic costs $L$ and $R$ are then used for regularization, for prioritizing attitude error reduction over velocity and general tuning ensuring smoothness of the generated solution. 

For including the $\infty-$norm cost in our SCP tool, we perform an \textit{epigraph trick}~\cite{Boyd2004} to reformulate it using linear constraints and by minimizing a slack variable $\gamma_i \geq 0$ in each stage which constrains
\begin{equation}
    -\gamma_i \leq S^{\frac{1}{2}} x_i\leq \gamma_i
\end{equation}

Regarding the use of the CMGs, even if explicit singularity avoidance is not included as a constraint, the fact of penalizing actuator effort implicitly implies certain singularity treatment, in the sense that the optimum profile will maximize torque capacities of the cluster, which discourages singular states.
This fact encourages the cluster to be in suitable configurations to provide high torques. Nevertheless, optimum profiles can include singular stages, e.g.\ if this singularity is not affecting the direction of demanded torque at this instant. Note that depending on the CMG configuration, measures to avoid singular configurations can be easily included in formulation~\eqref{eq:joint}, i.e.\ via constraints on the gimbal angles to avoid parallel orientations in the momentum discs.

\begin{figure}
\centering
\includegraphics[height=60mm]{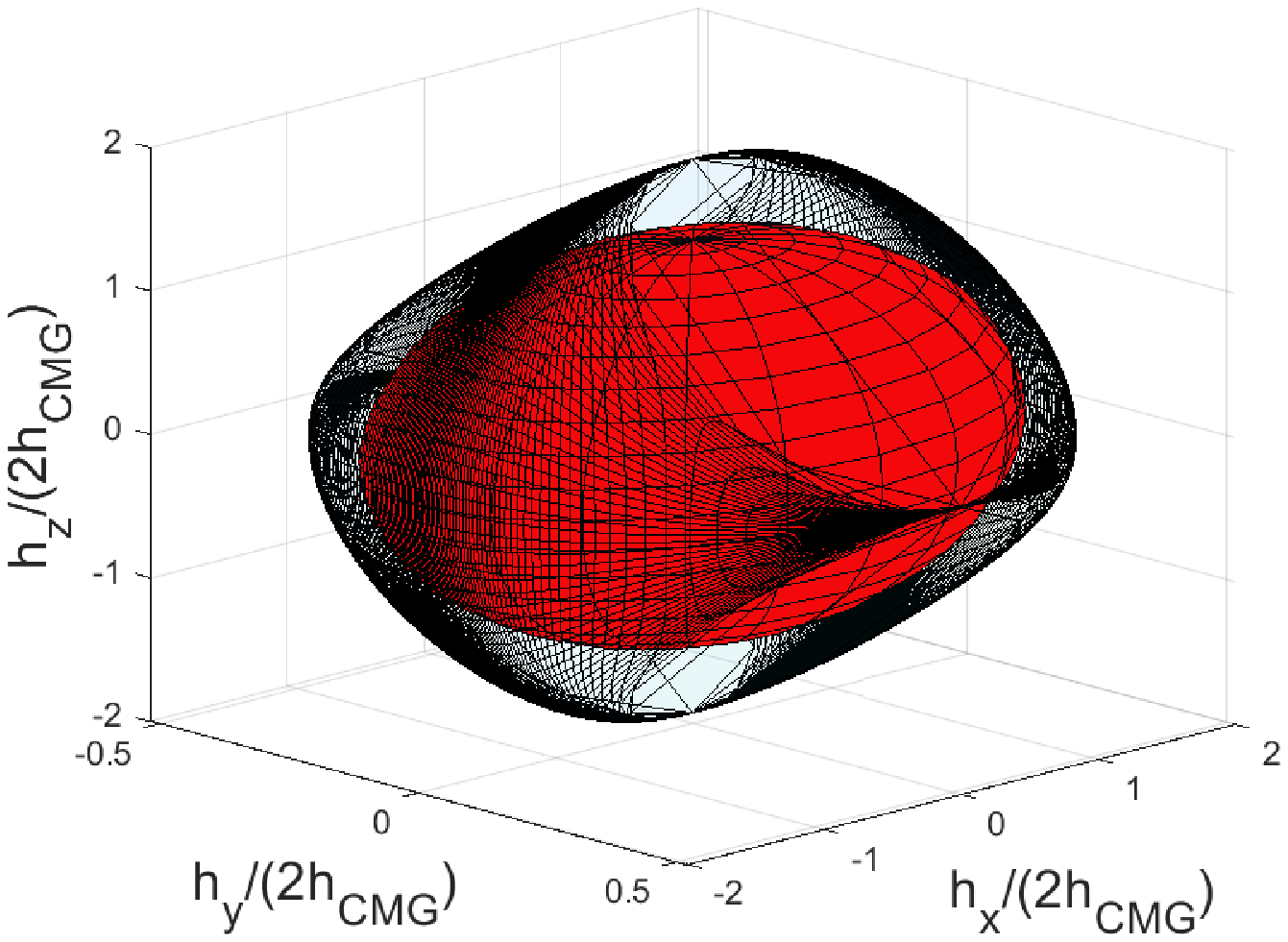}\includegraphics[height=60mm]{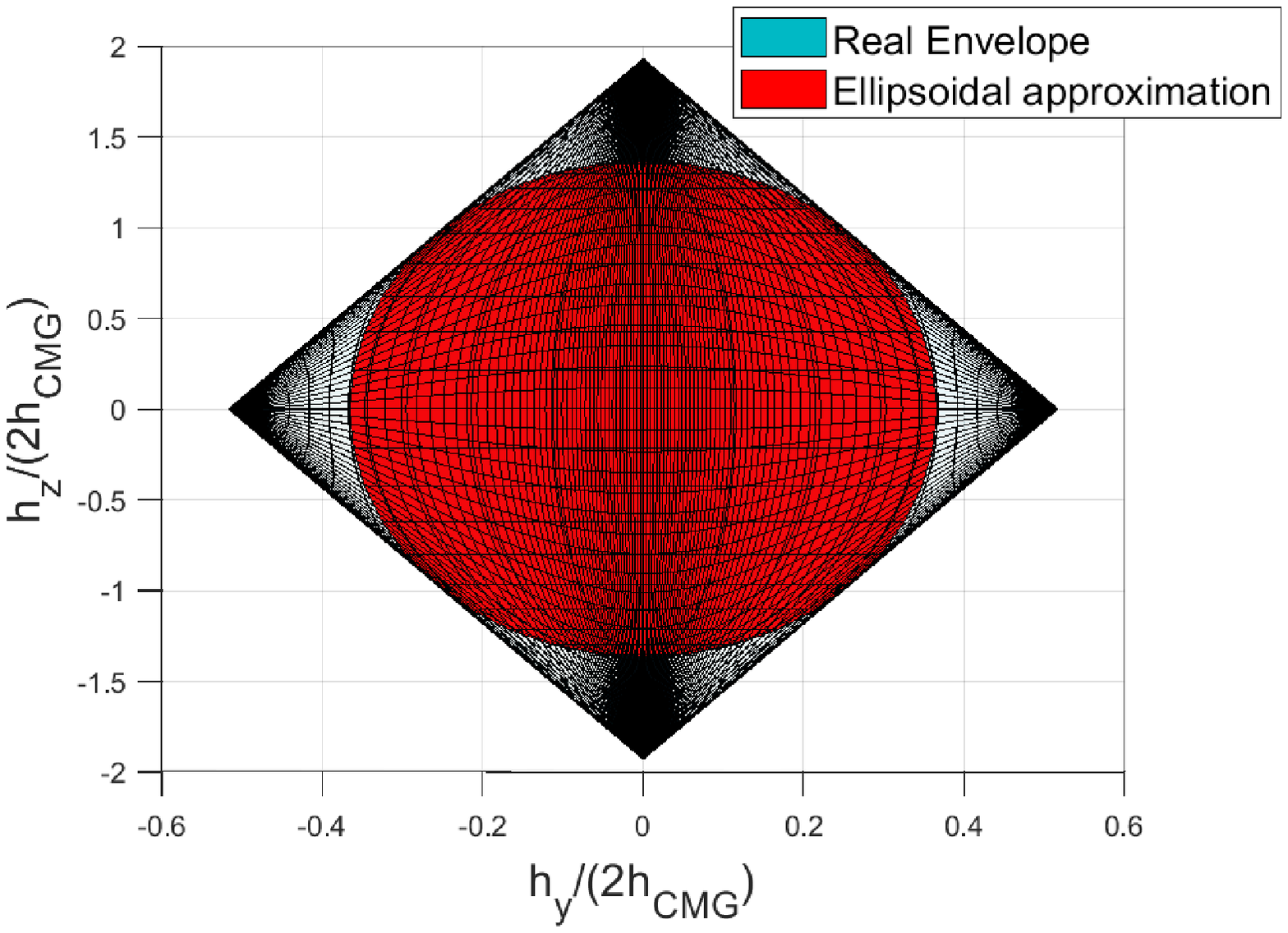}
\caption{Example of ellipsoidal approximation of roof CMG cluster with $\beta = 15$ degrees}
\label{fig:EllipsoidalConstraint}
\end{figure}

\subsubsection{Sequential optimization}
Since the joint optimization can be computationally heavy we investigate here a decomposition strategy following the usually split into attitude maneuver planning and CMG allocation. 

These two optimization problems are linked since the torque capabilities of the CMG cluster depend on the current gimbal angles. We, therefore, propose to divide the optimization into two different steps used in a sequential iterative process that decouples the highly non-linear dynamics of the CMGs from the attitude maneuver optimization. For this, we assume a CMG gimbal trajectory $\bar{\alpha}(t)$, either from the bang-bang solution (Section~\ref{ssc:Baseline}) or a previous iteration of the procedure, to formulate the reduced attitude dynamics 
\begin{equation} \label{eq:seq1dynamics}
\dot{x}=\begin{bmatrix} \begin{bmatrix}
\dot{\phi}\\
\dot{\theta}\\
\dot{\psi}
\end{bmatrix} \\
\dot{\omega}\\
\end{bmatrix}=
\begin{bmatrix}
(\omega_x\cos{\psi}-\omega_y\sin{\psi})\text{sec}\theta\\
\omega_x\sin{\psi}+\omega_y\cos{\psi} \\
\omega_z - (\omega_x\cos{\psi}-\omega_y\sin{\psi})\text{tan}\theta \\
-J^{-1}C_\tau \Lambda(\bar{\alpha}(t))\dot{\delta}\\
\end{bmatrix}
\end{equation}
In order to ensure that the outcome of this reduced optimization problem remains applicable to the overall system, we include two measures, namely
\begin{enumerate}
  \setlength\itemsep{0em}
    \item A trust-region like cost $||\dot{\delta}-\bar{\dot{\delta}}(t)\Vert_G^2$ on the gimbal rates to encourage the solution to remain close to the one used in the simplification step
    \item A quadratic momentum envelope constraint $\omega_i^\tp Q\omega_i \leq 1$ to take into account the capabilities of the CMG cluster. This is computed from the \textit{maximum volume ellipsoidal inner approximation of the Minkowski sum} \cite{Angulo2019} of the two CMG \textit{momentum discs}. The result of this approximation in the consider 4 CMG roof~\cite{Leve} case is shown in Figure \ref{fig:EllipsoidalConstraint}.
\end{enumerate}
Discretizing~\eqref{eq:seq1dynamics} again using a Runge Kutta scheme results in time-varying dynamics $f_i(x_i,u_i)$ allowing us to  formulate the attitude optimization problem as \\

{\medskip
\centering \emph{First: Attitude Optimization}
\begin{mini!}{x_i,\, u_i }{\frac{1}{2} ||x_{N}||_{L_f}^2+ ||x_N||_{S_f,\infty}
+
\sum_{i=0}^{N-1}  \frac{1}{2}||x_i||_L^2+ ||x_i||_{S,\infty}
+\frac{1}{2}||u_i\Vert_R^2
+\frac{1}{2}||u_i-\bar{\dot{\delta}}_{i}\Vert_G^2}{}{}
\addConstraint{x_{i+1}=f_i(x_i,u_i);\quad \forall i=0,..,N-1}
\addConstraint{\omega_i^\tp Q\omega_i \leq 1};\quad \forall i=1,...,N
\addConstraint{|u_i|\leq \dot{\delta}_{max};\quad \forall i=0,...,N-1}
\addConstraint{x_0 = x_{\text{init}}}
\end{mini!}}

With the solution of this previous optimization, a demanded momentum profile for the CMG cluster is generated $\bar{h}(t) =- J\bar{\omega}(t)$. This momentum profile is to be allocated in the gimbal rates, but no guarantees of singularity avoidance are provided, hindering direct allocation by classical steering laws~\cite{Leve}. Instead, a second optimization is carried out at the CMG level to minimize the error between momentum provided over time and demanded, while minimizing the actuation effort and thus managing singularities as discussed for the joint optimization. The outcome is therefore a dynamic CMG allocation, i.e.\ a reference trajectory of CMG gimbal angles. With $x = \alpha$ representing the gimbal angles, $u = \dot\delta$, and dynamics given via~\eqref{AlphaChangeVble2}, the CMG allocation optimization is expressed as \\

{\medskip \centering \emph{Second: CMG optimum allocation}
  \begin{mini!}{x_i, \, u_i }{\frac{1}{2} ||C_Hx_i+J\bar{\omega}_i||_{L_f}^2
+
\sum_{i=0}^{N-1}  \frac{1}{2}||C_Hx_i+J\bar{\omega}_i||_L^2
+\frac{1}{2}||u_i\Vert_R^2
+\frac{1}{2}||u_i-\bar{\dot{\delta}}_{i}\Vert_G^2}{}{}
\addConstraint{x_{i+1}=f_i(x_i,u_i);\quad \forall i=0,..,N-1}
\addConstraint{|u_i|\leq \dot{\delta}_{max};\quad \forall i=0,...,N-1}
\addConstraint{x_0 = x_{\text{init}}}
\end{mini!}}
Note that the term $\left(C_Hx_i+J\omega_i\right)$ is the error between commanded and CMG cluster momentum profiles.
We encapsulate the two optimizations in an iterative process in which, after the allocation, $\Lambda(\bar{\alpha}(t))$ is re-computed and the sequence repeated until convergence is achieved.

\subsection{Results and discussion}
\begin{figure}[h]
\begin{center}
\input{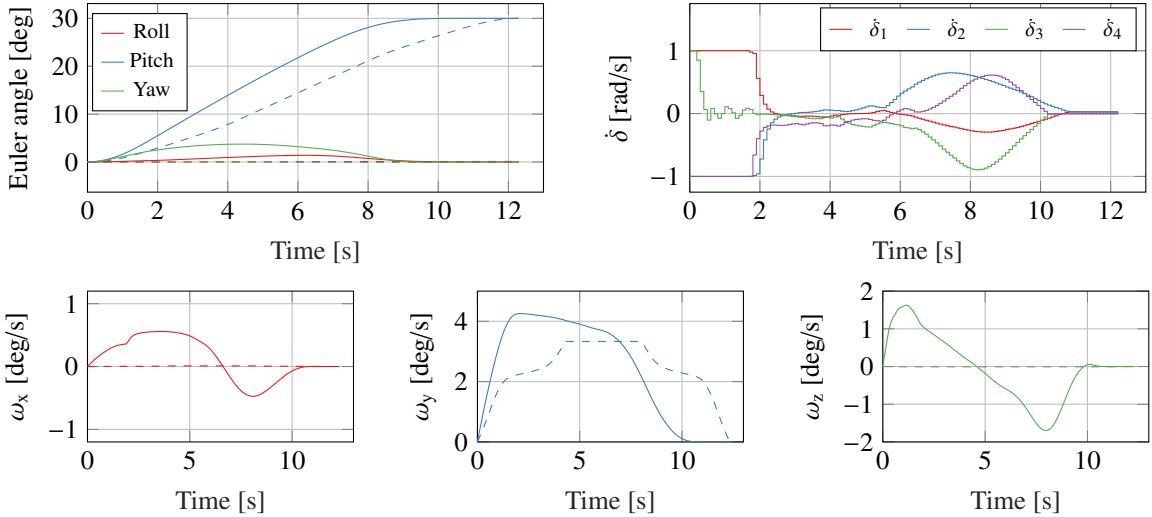}
\end{center} \vspace{-0.4cm}
\caption{Joint optimization VS baseline (dashed) for $\Delta \theta = 30$ degrees slew} \label{fg:joint_result}
\end{figure}

\begin{figure}
\begin{center}
\input{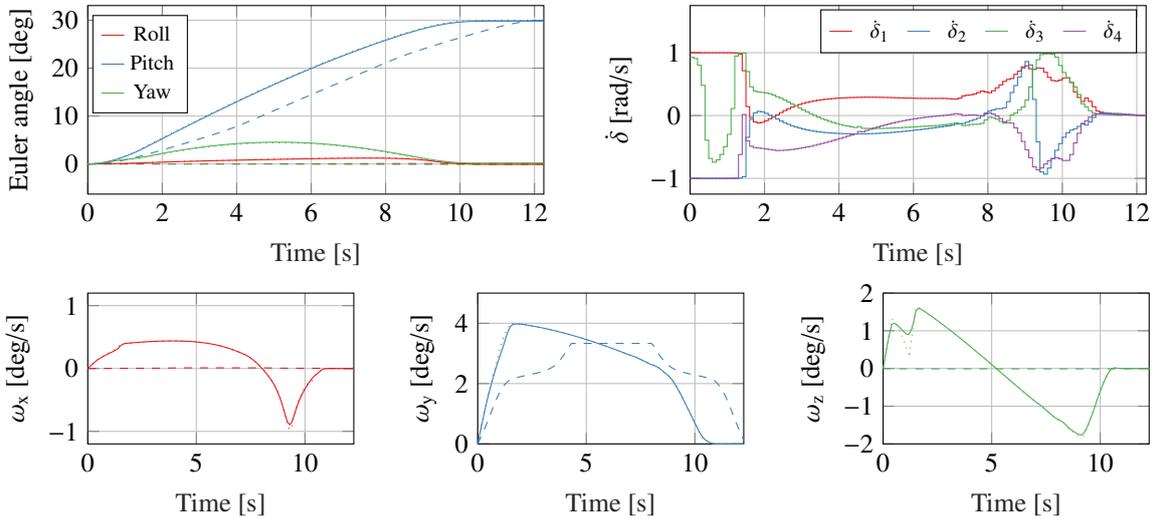}
\end{center} \vspace{-0.4cm}
\caption{Sequential optimization vs.\ baseline \emph{(dashed)} for $\Delta \theta = 30$ degrees slew. First (attitude) optimization represented with dotted lines. }
\label{fg:sequential_result}
\end{figure}
We test the developed guidance solutions for a scenario based on an agile missions study using CMGs reported in \cite{Geshnizjani2020}. For comparative purposes and optimization improvements demonstration, we perform a similar maneuver consisting of a simple slew around a pitch of $30$ degrees. We assume zero velocities of the spacecraft at the beginning of the maneuver and a suitable gimbal configuration with:
\begin{equation}
    \delta_1 =\frac{\pi}{3};\quad \delta_2 = -\frac{\pi}{3} ;\quad \delta_3 = \pi-\frac{\pi}{3} ;\delta_4 = \pi+\frac{\pi}{3} ;
\end{equation}

For the optimization, we discretize continuous-time dynamics using a Runge Kutta 4 scheme with a discretization time of $T_s = 0.1\,\text{s}$ and estimate the time horizon $N$ based on the bang-bang-like baseline solution which we also use to warm-start all optimization formulations.

Figure~\ref{fg:joint_result} shows the outcome of the joint optimization of CMG and attitude trajectories. As can be seen, this is able to reduce the maneuver time by about $12.2\%$ (2 seconds) to a steady state compared to the baseline solution. It can be seen that the optimization exploits nondiagonal inertia properties, as well as the CMG torque capabilities in a non-trivial way -- see e.g.\ how 3 of the 4 CMGs act in saturation during the beginning of the maneuver. Even if there are no appreciable errors on the guidance profiles, some small target offsets appear in a non-linear propagation of the gimbal rate profiles, e.g.\ a residual velocity $\left(<0.005\text{ deg/s}\right)$ and attitude error $\left(<0.4\text{ deg}\right)$. Computationally the formulation presents challenges due to the highly nonlinear CMG dynamics interplaying with the attitude dynamics which seem to prohibit rapid convergence.

This situation is improved using the presented sequential scheme, which reliably finds feasible solutions after just two sequence steps consisting of attitude and CMG optimization, which can result in a speedup by an order of magnitude (depending on the precision required by the joint optimization). Figure~\ref{fg:sequential_result} shows the outcome of this procedure, again leading to a maneuver time reduction of about 11.4~\% compared to the baseline solution. While the resulting profiles are quite distinct from the joint optimization, they share that they exploit the torque capacities of the CMG array resulting in saturation of 3 of the 4 CMGs at the beginning of the maneuver. Again, we have small errors in the nonlinear propagation in the terminal velocities and attitudes with residual velocities below $0.004$ deg/s and attitude error $<0.2$ degrees. Again these are largely due to discretization errors, but in this case also accumulate from momentum tracking errors in the CMG optimization.

\begin{figure}
\begin{center}
\input{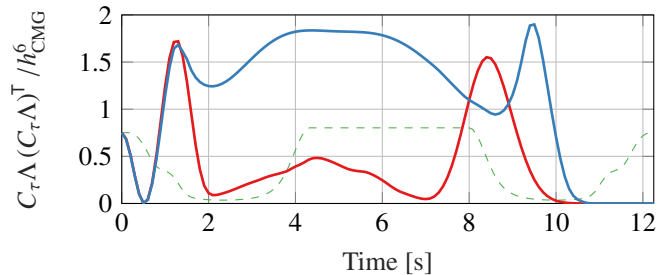}
\end{center}
\vspace{-0.4cm}
\caption{Determinant of the CMG Jacobian as a singularity metric over the executed slew maneuver. \emph{(Dashed green)} Baseline bang-bang profile. \emph{(red)} Joint optimization. \emph{(blue)} Sequential optimization.}
\label{fig:Singularity}
\end{figure}

Regarding the CMG allocation and singularity management, it can be appreciated in Figure \ref{fig:Singularity} how the cluster gets prepared at the times of high torque demanded to be far from singularity, so that more actuation is available. This effect can be especially appreciated at the brake point at around 10 seconds. As expected, singularity is not avoided but rather managed so as to minimize the effect on overall performance. Differences exist between joint and sequential solutions, mainly, since the joint optimization is able to exploit the full actual momentum envelope of Figure \ref{fig:EllipsoidalConstraint}, and the sequential one is sensitive to the warm starting using $\Lambda(\bar{\alpha}(t))$. The final singular situation when the slew has ended is due to the actuation minimization. i.e., the cluster is taken to zero momentum with minimum effort. Note that the formulation of the optimization problems can be easily adapted, e.g.\ to include preferred gimbal angles, to prepare the initial configuration for the next maneuver, or even include some \textit{preparation time} as part of the optimization so that it autonomously prepare the cluster in an optimum initial configuration for next slew. 

\section{Conclusion}\label{sc:conclusion}
This paper has outlined the developments of the SOTB, a SENER Matlab toolbox for numerical optimization and control, highlighting its capabilities with regard to convex as well as nonlinear problems. 
We have demonstrated the potential of such numerical optimization in agile attitude maneuvers under CMG actuation. In particular, the optimization schemes proposed manage to reduce maneuvers time up to $12.2\%$ while generating smooth profiles and automatically dealing with CMG cluster singularities. In this matter, singularities are not completely avoided, as in classical steering laws, but rather \textit{traversed} in a way so that the complete slew profile is optimum in time and control required. The so-called \textit{sequential optimization} formulation has demonstrated a major improvement in terms of computation time at a slight reduction in performance w.r.t.\ the \textit{joint optimization} -- while both show significant performance improvement over a classical \emph{bang-bang}-like strategy. We have furthermore highlighted the flexibility in optimization-based formulations, which allow for multitude of adaptations and extensions to different scenarios.

Future work will require extensive testing of the generated guidance profiles in a realistic simulator accounting for complete dynamics, perturbations, and navigation uncertainties. In addition, we plan to implement the developed algorithms on space-graded hardware to assess computation times and feasibility under restricted resources. 

\FloatBarrier
\bibliography{bibliography}
\bibliographystyle{unsrt}

\end{document}